\documentclass[10pt,conference]{IEEEtran}
\usepackage{cite}

\ifCLASSINFOpdf
\else
\fi
\usepackage{amsmath,amssymb,amsthm,amsfonts,amscd,amsbsy,amsxtra,amsfonts}
\usepackage{mathtools}
\DeclareMathAlphabet{\mathpzc}{OT1}{pzc}{m}{it} 
\usepackage{url}

\usepackage[dvipsnames]{xcolor}
\usepackage[normalem]{ulem}
\usepackage{array}
\usepackage{makecell}

\DeclareMathOperator{\sinc}{Sinc}
\DeclareMathOperator{\rect}{\Pi}

\newtheorem{lemma}{\underline{Lemma}}

\begin{document}
%
\title{On Delay-Doppler Plane Orthogonal Pulse}
%
%
%

\author{\IEEEauthorblockN{Hai Lin}
  \IEEEauthorblockA{Osaka Metropolitan University\\
    Sakai, Osaka, 599-8531, Japan\\
    Email: hai.lin@ieee.org}\\\vspace*{-1.007cm}
  \and
  \IEEEauthorblockN{Jinhong Yuan}
  \IEEEauthorblockA{The University of New South Wales\\
    Sydney, NSW, 2052, Australia\\
    Email: j.yuan@unsw.edu.au}\\\vspace*{-1.007cm}
}


%
%

\markboth{Journal of \LaTeX\ Class Files,~Vol.~14, No.~8, August~2015}%
{Shell \MakeLowercase{\textit{et al.}}: Bare Demo of IEEEtran.cls for IEEE Journals}
%



\maketitle

\begin{abstract}In this paper, we analyze the recently discovered delay-Doppler plane orthogonal pulse (DDOP), which is essential for delay-Doppler plane multi-carrier modulation waveform.
  In particular, we introduce a \emph{local orthogonality} property of pulses corresponding to Weyl-Heisenberg (WH) \emph{subset} and
  justify the DDOP's existence,
  in contrast to \emph{global orthogonality} corresponding to WH \emph{set} governed by the WH frame theory. Then, sufficient conditions for locally-orthogonal pulses are presented and discussed. Based on the analysis, we propose a general DDOP design. We also derive the frequency domain representation of the DDOP, and compare the DDOP-based orthogonal delay-Doppler division multiplexing (ODDM) modulation with other modulation schemes, in terms of TF signal localization. Interestingly, we show  perfect local orthogonality property
  of the DDOP with respect to delay-Doppler resolutions using its ambiguity
  function.
\end{abstract}


%
\IEEEpeerreviewmaketitle

\section{Introduction}
%
%
%
%
In digital communications, a modulation scheme usually requires a set of
(bi)orthogonal \emph{analog} pulses or continuous time functions,
each of which carries an information-bearing \emph{digital} symbol, to synthesize the signal waveform\cite{Jacobs_pce}. Therefore, the modulation process can be intuitively thought of as placing these pulses in the time-frequency (TF) plane, and the (bi)orthogonality can be achieved by placing them with proper TF distance.
Such modulation schemes include single carrier (SC) modulation with temporally spaced pulses, and multi-carrier (MC) modulation whose pulses are spaced both temporally and spectrally.
Meanwhile, for a communication system,
a transmit signal always consists of a finite number of pulses and occupies a finite TF region in the TF plane, determining the signal's duration and bandwidth.

In the context of MC modulation, the pulses are typically generated by TF shifting a \emph{prototype pulse} in accordance with a frequency resolution $\mathcal F$
and a time resolution $\mathcal T$.
The minimum TF distance among these pulses can be quantified by $\mathcal R= \mathcal T \mathcal F$, called as joint TF resolution (JTFR) in this paper. The fundamental issue of designing an MC modulation scheme is to find the prototype pulse that can form (bi)orthogonal pulses with respect to $\mathcal T$ and $\mathcal F$.
Conventionally, these TF-shifted pulses are considered as \emph{Wely-Heisenberg} (WH) or \emph{Gabor} function set\cite{wavelet,gaborana,ftfa}.
According to the WH frame theory, the (bi)orthogonal WH function sets only exist for $\mathcal R \ge 1$\cite{wexler1990,janssen1995}, and therefore most of orthogonal MC modulation schemes are designed with $\mathcal R \ge 1$ \cite{tff,mct}.

Recently, a delay-Doppler plane MC (DDMC) modulation named as the orthogonal delay-Doppler division multiplexing (ODDM) modulation, was proposed in \cite{oddmicc22,oddm}. Considering that linear time-varying (LTV) channels in a stationary region that can be modelled as a delay-Doppler (DD) channel with a deterministic spreading function, the ODDM modulation employs a newly discovered DD plane orthogonal pulse (DDOP) to couple the modulated MC signal with the DD channel. It achieves superior performance by harvesting both time and frequency diversity, while
it is shown in \cite{oddmicc22,oddm} that
the DDOP can form an orthogonal function set with respect to the DD plane resolutions.
Because the DD plane's TF resolutions result in a JTFR $\mathcal R_{\textrm{DD}} < 1$, the DDOP seems inconsistent with current (bi)orthogonal pulses design principles.
Although its orthogonality has been proved, a rational explanation for the DDOP's unique properties is still missing.

In this paper, we take an in-depth look into the DDOP and justify its existence.
We introduce a \emph{local orthogonality} property and  clarify that the DDOP only needs to satisfy local orthogonality,
in contrast to \emph{global orthogonality} governed by the WH frame theory.
Then, sufficient conditions for pulse to achieve local orthogonality are analyzed. Based on the analysis, we propose a general DDOP design.
Our contributions can be summarized as follows:
\vspace{-0.5mm}
\begin{itemize}

  \item We point out that only local (bi)orthogonality in the finite TF region rather than global (bi)orthogonality in the whole TF plane is required by a modulation scheme. Accordingly, we show that a WH \emph{subset} rather than a WH set is required in the pulse design.

  \item We reformulate the (bi)orthogonal pulse design problem, based on the local (bi)orthogonality. We show that the DDOP forms a WH subset that satisfies the local orthogonality.

  \item We analyze the local orthogonality with respect to TF resolutions, and discuss the corresponding sufficient conditions. We reveal that for a limited number of subcarriers, surprisingly, there are \emph{infinite} pulses orthogonal with respect to $\mathcal F$, as long as they are periodic functions with a specified period related to the number of subcarriers.

  \item By introducing cyclic prefix (CP) and cyclic suffix (CS) to achieve the specified periodicity, we propose a general DDOP design, which releases the duration constraint of square-root Nyquist (SRN) sub-pulses in our previously designed DDOP.

  \item We derive the frequency domain representation of the DDOP. Together with the DDOP's time domain representation, we illustrate the DDOP-based ODDM's TF signal localization, and
        schematically compare it with those of other modulation schemes.
        The ambiguity function shows perfect local orthogonality property of the DDOP with respect
        to delay-Doppler resolutions.

\end{itemize}

Notations: In this paper,
$\rect_{\mathfrak T}(t)$ stands for the rectangular pulse with unit energy and support $[0, \mathfrak T]$. Given the number of subcarriers $N$, $a(t)$ denotes the SRN pulse for interval $\mathcal T$, with energy $\frac{1}{N}$ and support $[-T_a/2, T_a/2]$. $\mathcal A_{g,\gamma}(\cdot)$ is the (cross)ambiguity function of $g(t)$ and $\gamma(t)$.

\section{WH set based pulse design principles}
Let us first introduce main parameters and their notations for an MC modulation in Table \ref{tab:parameters}.
The transmit pulses in an MC modulation
can be represented by the function set
\begin{equation}\label{gtf}
  \left(g,\mathcal T, \mathcal F\right)=\left\{g_{m,n}\right\}_{m,n\in \mathbb Z},
\end{equation}
where $g_{m,n}\coloneqq g(t-m\mathcal T)e^{j2\pi n \mathcal F (t-m\mathcal T)}$ and $g(t)$ is the prototype pulse.
Similarly, we can form the receive pulses $\left(\gamma,\mathcal T, \mathcal F\right)$ using another prototype pulse $\gamma(t)$ with the same TF resolutions.
Note that because a time-limited signal cannot be strictly band-limited, the bandwidth of $g(t)$, $B_g$, is defined in an essential sense \cite{onbandwidth}.
\begin{table}[t]
  \centering
  \caption{MC Modulation Parameters }
  \bgroup
  \def\arraystretch{1.2}
  \begin{tabular}{|c|c|}
    \hline
    \thead{Notation} & \thead{Parameter}                                                  \\
    \hline
    $\mathcal F$     & \makecell{frequency resolution,                 frequency spacing} \\ 
    \hline
    $\mathfrak T$    & symbol period, $\mathfrak T=1/\mathcal F$                          \\
    \hline
    $\mathcal T$     & time resolution, symbol interval                                   \\
    \hline
    $\mathcal R$     & JTFR, $\mathcal R=\mathcal T\mathcal F$                            \\
    \hline
    $N$              & number of subcarriers                                              \\
    \hline
    $M$              & number of symbols                                                  \\
    \hline
    $g(t)$           & transmit (prototype) pulse                                         \\
    \hline
    $T_g$            & duration of $g(t)$, symbol duration                                \\
    \hline
    $B_g$            & bandwidth of $g(t)$                                                \\
    \hline
  \end{tabular}
  \egroup
  \label{tab:parameters}
\end{table}

Given $\mathcal T$ and $\mathcal F$, the fundamental issue of an MC modulation is to find $g(t)$ and $\gamma(t)$ satisfying the orthogonal condition of
\begin{equation}\label{g_ortho}
  \langle g_{m,n}, g_{\dot m,\dot n}\rangle =   \delta(m-\dot m)\delta( n-\dot n),
\end{equation}
or the biorthogonal condition of
\begin{equation}\label{g_biortho}
  \langle g_{m,n}, \gamma_{\dot m,\dot n}\rangle =\delta(m-\dot m)\delta(n-\dot n).
\end{equation}
By considering the TF plane as a 2D phase space, the function set in (\ref{gtf}) forms a discrete lattice ``sampling" the phase space\cite{haas_wpc_97,wavelet}, where the ``sampling" resolution is the JTFR $\mathcal R$.
Then, the function set in (\ref{gtf}) can be treated as a WH set.
According to the WH frame theory, the existence of (bi)orthogonal WH set depends on the ``sampling" resolution and can be summarized as \cite{cofdm,wavelet,ftfa,gaborana,wexler1990,haas_wpc_97,kozek98,tff,strohmer2003}:
\begin{itemize}
  \item Critical sampling ($\mathcal R =1$) : Orthogonal WH sets exist. However, they have either infinite time or frequency energy spread according to the Balian-Low theory \cite{wavelet_book}, and therefore are not TF well-localized.
  \item Undercritical sampling ($\mathcal R >1$) : TF well-localized orthogonal or biorthogonal WH sets exist, if $\mathcal R$ is sufficiently larger than $1$.
  \item Overcritical sampling ($\mathcal R <1$) : Neither orthogonal nor biorthogonal WH sets exist.
\end{itemize}

With the transmit pulses in (\ref{gtf}), the transmit waveform of an MC modulation can be represented as
\begin{align}\label{xt}
  x(t)=\sum_{m=0}^{M-1}\sum_{n=0}^{N-1} X_{m,n}g(t-m\mathcal T)e^{j2\pi n \mathcal F (t-m\mathcal T)},
\end{align}
where $X_{m,n}$'s are the information-bearing digital symbols.

\section{ODDM modulation}
In the design of modulation schemes, the primary concern is the dispersive effect of the channel.
A doubly-selective wireless channel with both time and frequency dispersion is usually considered as a LTV system, and represented by its time-varying channel impulse response (TV-CIR) or DD spread function\cite{bello}.

\subsection{DD channel model}

Since the transmit signal is band- and time-limited, we always apply an appropriate bandpass filtering and a subsequent sampling at the receiver. As a result, we observe an equivalent channel that is the band- and time-limited version of the physical channel.
Let the sampling rate and duration be $W_0$ and $T_0$, respectively. The equivalent DD channel
can be written as \cite{bello}
\begin{equation}\label{ddchannel}
  h(\tau,\nu)=\sum_{p=1}^{P}h_p \delta (\tau-\tau_p)\delta (\nu-\nu_p),
\end{equation}
with 
$\tau_p=\frac{l_p}{W_0}$, $\nu_p=\frac{k_p}{T_0}$, $l_p, k_p \in \mathbb Z$, where $\frac{1}{W_0}$ and $\frac{1}{T_0}$ are the delay and Doppler resolutions, respectively.

\begin{figure}
  \centering
  \includegraphics[width=8.5cm]{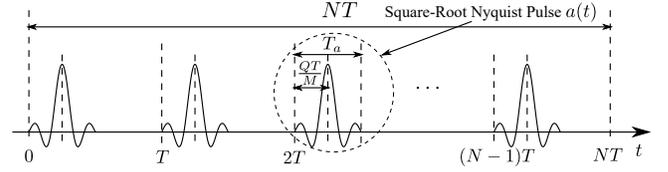}
  \caption{$u(t)$, the transmit pulse of ODDM modulation.}
  \label{shape_ut}
  \vspace{-5mm}
\end{figure}

\subsection{ODDM modulation and DDOP}
To couple the MC signal with the DD channel in
(\ref{ddchannel}), the ODDM matches its signal resolutions to the delay and Doppler resolutions, namely set $\mathcal T=\frac{1}{W_0}$ and $\mathcal F=\frac{1}{T_0}$, respectively. Note that for an ODDM signal, we have $W_0=\frac{M}{T}$ and $T_0=NT$. Then, an ODDM frame without the frame-wise CP can be written as\cite{oddm}
\begin{align}\label{xtoddm}
  x(t)=\sum_{m=0}^{M-1}\sum_{n=0}^{N-1} X_{m,n}u\left(t-m\frac{T}{M}\right)e^{j2\pi n \frac{1}{NT} (t-m\frac{T}{M})},
\end{align}
where $u(t)$ is the DDOP given by
\begin{equation}\label{ut}
  u(t)=\sum_{\dot n=0}^{N-1}a(t-\dot nT).
\end{equation}
As shown in Fig. \ref{shape_ut}, the duration of
$a(t)$ in $u(t)$ is $T_a=2Q\frac{T}{M}$.
When $2Q\ll M$ and therefore $T_a\ll T$, it has been proved in \cite{oddm} that $u(t)$ satisfies the orthogonal property
\begin{equation}
  \mathcal A_{u,u}\left(\bar m\frac{T}{M}, \bar n\frac{1}{NT}\right)= \delta(\bar m)\delta(\bar n),\label{orth}
\end{equation}
for $|\bar m|\le M-1$ and $|\bar n| \le N-1$.
Because the corresponding JTFR of $\mathcal R_{\textrm{DD}}=  \frac{T}{M}\times \frac{1}{NT}=\frac{1}{MN}\ll 1$ does not allow the existence of (bi)orthogonal WH set,
a natural question arises: How to explain the existing DDOP in \cite{oddm} and whether is there any general DDOP design principle?

\section{Global and Local (Bi)Orthogonality}

From (\ref{orth}), one can see that this orthogonality is regarding $M$ symbols with $N$ subcarriers, and therefore {it only applies to a part of TF plane}.
{Since} an MC modulation has {a} limited number of symbols and subcarriers,  the orthogonality {within this signal bandwidth and duration} is sufficient for an MC modulation. As a result, we can reformulate its pulse design problem, {and introduce a concept of} local orthogonality.

\subsection{Global and local (bi)orthogonality}
Analogous to (\ref{g_ortho}) and (\ref{g_biortho}), the (bi)orthogonal pulse design problem taking the limited number of symbols and subcarriers
into account is
to find WH subsets $ \left(g,\mathcal T, \mathcal F, M, N\right)$ and $ \left(\gamma ,\mathcal T, \mathcal F, M, N\right)$ that satisfy the orthogonal condition of
\begin{equation}\label{l_ortho}
  \langle g_{m,n}, g_{\dot m,\dot n}\rangle =\delta(m-\dot m)\delta(n-\dot n),\,\, m,\dot m\in \mathbb Z_m, n,\dot n\in \mathbb Z_N,
\end{equation}
or the biorthogonal condition of
\begin{equation}\label{l_biortho}
  \langle g_{m,n}, \gamma_{\dot m,\dot n}\rangle =\delta(m-\dot m)\delta(n-\dot n),\,\, m,\dot m\in \mathbb Z_m, n,\dot n\in \mathbb Z_N,
\end{equation}
where
\begin{align}
  \mathbb Z_M =\{0, 1, \cdots, M-1\}, \,\, \mathbb Z_N =\{0, 1, \cdots, N-1\}.
\end{align}
We call (\ref{l_ortho}) and (\ref{l_biortho}) the local orthogonal
condition and local biorthogonal condition, respectively.
Because of
\begin{align}
  \langle g_{m,n}, g_{\dot m,\dot n}\rangle
  = \mathcal A_{g,g} (\bar m \mathcal T, \bar n \mathcal F)e^{j2\pi n\bar m \mathcal F \mathcal T},
\end{align}
where $\bar m= \dot m-m$ and $\bar n= \dot n-n$, the local orthogonal condition in (\ref{l_ortho}) is equivalent to
\begin{align}\label{agg}
  \mathcal A_{g,g}(\bar m\mathcal T, \bar n\mathcal F)=\delta(\bar m)\delta(\bar n),
\end{align}
for $|\bar m| \le M-1, |\bar n| \le N-1$. Similar result can be obtained for the local biorthogonal condition in (\ref{l_biortho}). 

It is noteworthy that the WH frame theory based results regarding (bi)orthogonal WH sets are rigorously correct. Since the WH set is a time-frequency analysis tool for functions in $L^2(\mathbb R)$, it considers the whole TF plane where $m,n\in \mathbb Z$, and corresponds to the signal without the limitation of bandwidth and duration.
To make this possible, given $\mathcal T$ and $\mathcal F$, $g(t)$ is independent of the number of symbols $M$ and the number of subcarriers $N$, to be shifted over the whole TF plane. In other words, to achieve the global (bi)orthogonality in (\ref{g_ortho}) and (\ref{g_biortho}), $g(t)$ is parameterized  only by  $\mathcal T$ and/or $\mathcal F$.

On the other hand, for MC modulation,
a WH subset that satisfies the local (bi)orthogonality in (\ref{l_ortho}) and (\ref{l_biortho}) is sufficient.
Obviously, $g(t)$ that achieves the global (bi)orthogonality can form a such WH subset.
However, what we really need is just a WH subset, and it is not necessarily bounded by the WH frame theory for the WH set.
In fact, the pulses parameterized by not only $\mathcal T$ and $\mathcal F$ but also $M$ and $N$, can achieve the local orthogonality.
An example is the DDOP in (\ref{ut}).

\subsection{Orthogonality with respect to $\mathcal F$}

\begin{figure}
  \centering
  \includegraphics[width=7cm]{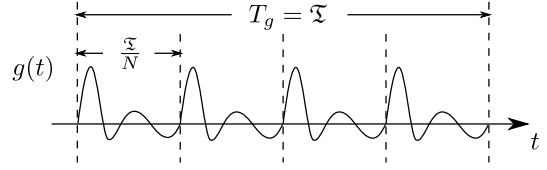}
  \caption{$g(t)$ orthogonal w.r.t $\mathcal F=\frac{1}{\mathfrak T}$ for $|n|\le N-1$ and fixed $m$.}
  \label{gtF2}
  \vspace{-5mm}
\end{figure}

Let us consider a fixed $m$ in $g_{m,n}$, and investigate the orthogonality with respect to the frequency resolution $\mathcal F$. We want to find $g(t)$ that can achieve the orthogonality among $g(t-m\mathcal T)e^{j2\pi n \mathcal F (t-m\mathcal T)}$ with a given $m$ but variable $n$, where $0\le  t \le T_g$ and $T_g=\mathfrak T=1/\mathcal F$. Without loss of generality, let $m=0$. We can obtain the following results:
\renewcommand{\labelenumi}{(F\arabic{enumi})}
\begin{enumerate}
  \item Unbounded $n$ ($n\in \mathbb Z$):
        $g(t)$ is the rectangular pulse $\Pi_{\mathfrak T}(t)$, which is independent of $N$.

  \item Bounded $n$ ($|n|\le N-1$): We have the following lemma:
        \begin{lemma}\label{l1}
          When $g(t)$ is a periodic function with period $\frac{\mathfrak T}{N}$ for $0\le  t \le T_g$ and $T_g=\mathfrak T$, it satisfies the orthogonal property that
          \begin{equation}
            \mathcal A_{g,g}\left(0, n\mathcal F\right)
            =\delta(n),
          \end{equation}
          for $|n|\le N-1$.
        \end{lemma}
        \begin{IEEEproof}
          Since the period of $g(t)$ is $\frac{\mathfrak T}{N}$, $g(t)$ can be written as
          \begin{align}
            g(t)=g\left(t+\dot n\frac{\mathfrak T}{N}\right), \,\,\,0 \le \dot n \le N-1.
          \end{align}
          for $0\le  t < \frac{\mathfrak T}{N}$.
          Then, bearing in mind that $\mathfrak T=1/\mathcal F$, we have 
          \begin{align}
             & \mathcal A_{g,g}(0,n\mathcal F)
            \nonumber                                                                                                                                          \\
             & =  \int_0^{T_g} g(t)g^*(t)e^{-j2\pi n\mathcal Ft} dt, \nonumber                                                                                 \\
             & = \sum_{\dot n=0}^{N-1} \int_{\dot n \frac{\mathfrak T}{N}} ^{(\dot n+1) \frac{\mathfrak T}{N}} g(t)g^*(t)e^{-j2\pi n\mathcal Ft} dt, \nonumber \\
             & = \sum_{\dot n=0}^{N-1} e^{-j2\pi\frac{n\dot n}{N}}  \int_0^{\frac{\mathfrak T}{N}} g(t)g^*(t)e^{-j2\pi n\mathcal Ft} dt, \nonumber             \\ \label{northo}
             & = \delta(n),
          \end{align}
          for $|n|\le N-1$, which completes the proof.
        \end{IEEEproof}
\end{enumerate}
Lemma \ref{l1} indicates that once there is a constraint imposed on the number of subcarriers, there are \emph{infinite} pulses that can satisfy the orthogonality with respect to $\mathcal F$.   In particular, \emph{regardless of $B_g$}, $g(t)$
can achieve the orthogonality among $N$ subcarriers with a subcarrier spacing $\mathcal F$, as long as it is an aforementioned periodic function. An example of such $g(t)$ for $N=4$ is shown in Fig. \ref{gtF2}.

It is noteworthy that in contrast to (F1) where $B_g$ is proportional to $\mathcal F$, (F2) \emph{decouples} $B_g$ and $\mathcal F$, and consequently allows pulses with much wider bandwidth to achieve orthogonality among $N$ subcarriers. On the other hand, to avoid the intersymbol interference (ISI) and achieve the orthogonality among MC symbols time-multiplexed by $\mathcal T$, we need $B_g$ to be comparable to $\frac{1}{\mathcal T}$.
The decoupling of $\mathcal F$ and $B_g$ in (F2) actually paves a way to design orthogonal pulse with respect to \emph{independent TF resolutions}.

\subsection{Orthogonality with respect to $\mathcal T$}
Similarly, we can consider a fixed $n$ in $g_{m,n}$, and investigate the orthogonality with respect to the time resolution $\mathcal T$.
Our target now is to find $g(t)$ that can achieve the orthogonality among $g(t-m\mathcal T)e^{j2\pi n \mathcal F (t-m\mathcal T)}$ with a fixed $n$ but different $m$. When $n\ne 0$, we have the following straightforward answer with  \emph{isolated} pulses/sub-pulses:
\renewcommand{\labelenumi}{(T\arabic{enumi})}
\begin{enumerate}
  \item Unbounded $m$ ($m\in \mathbb Z$) : $g(t)$ can be any function with duration $T_g\le \mathcal T$, which is independent of $M$.
  \item Bounded $m$ ($|m|\le M-1$) : $g(t)$ consists of $\dot N >1 $ sub-pulses $b_{\dot n}(t), 0\le \dot n \le N-1$, where these sub-pulses are temporally spaced by $M\mathcal T$ and each sub-pulse has a duration of $T_{b_{\dot n}} \le \mathcal T$.
\end{enumerate}
Meanwhile, when $n=0$, we have another answer with \emph{overlapped} pulse/sub-pulses:
\begin{enumerate}
  \setcounter{enumi}{2}
  \item Unbounded $m$ ($m\in \mathbb Z$) : SRN pulse for symbol interval $\mathcal T$, which is also independent of $M$.
  \item Bounded $m$ ($|m|\le M-1$) : $g(t)$ consists of $\dot N >1 $ SRN sub-pulses for symbol interval $\mathcal T$,  where these sub-pulses are temporally spaced by $M\mathcal T$. The SRN sub-pulse can have any duration.
\end{enumerate}
It is interesting to note that $g(t)$ in (T4) actually can form a periodic function that satisfies (F2), when $\dot N$ is large enough.

\begin{figure}
  \centering
  \includegraphics[width=9cm]{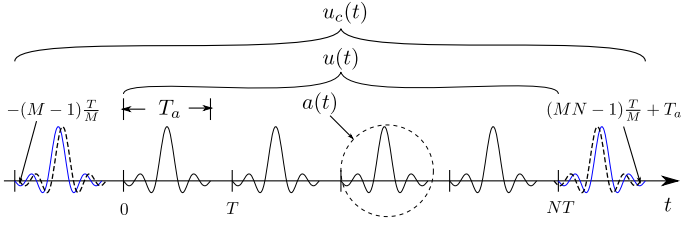}
  \caption{$u_c(t)$ for $D=1$.}
  \label{gce1}
  \vspace{-5mm}
\end{figure}

\section{General DDOP design}
Recall that the orthogonal property of the DDOP in (\ref{orth}) is subject to a duration constraint of SRN sub-pulse, given by $T_a\ll T$. In practice, it is desirable to relax such constraint to enable flexible design. In this section, we propose a general DDOP design, where the SRN sub-pulse's duration constraint is released.

Let $\dot N=N$ and $\mathcal T=\frac{T}{M}$, $g(t)$ in (T4) becomes the DDOP $u(t)$ in (\ref{orth}), except the unbounded $T_a$. From (F2), we know that for the frequency resolution $\mathcal F=\frac{1}{NT}$, the key to achieve the orthogonality among $N$ subcarriers is to form a periodic function with period $\frac{1}{N\mathcal F}=T$. This observation inspires us to use $u_c(t)$, a cyclically extended version of $u(t)$, as the transmit pulse, while the receive pulse is still $u(t)$. Furthermore, because $\mathcal A_{u_c,u}(m\frac{T}{M},  n\frac{1}{NT})$ is calculated between $u_c(t)$ and $u(t- m\frac{T}{M})e^{j2\pi \frac{ n}{NT}(t- m\frac{T}{M})}$,
the problem becomes how can we let $u_c(t)$ have the specified periodicity within the range of $u(t-m\frac{T}{M})e^{j2\pi \frac{\bar n}{NT}(t-m\frac{T}{M})}$ for $|m|\le M-1$. We have the following lemma:

\begin{lemma}\label{l2}
  Let $u(t)$ consist of $N$ SRN pulses $a_{T/M,N}(t)$ temporally spaced by $T$, it satisfies the orthogonal property that
  \begin{equation}\label{afgceg}
    \mathcal A_{u_c,u}\left(m\frac{T}{M},n\frac{1}{NT}\right)= \delta(m)\delta(n),
  \end{equation}
  for $|m|\le M-1$ and $|n|\le N-1$, where
  $u_c(t)$ is a cyclically extended version of $u(t)$ that is a periodic function with period $T$ during $-(M-1)\frac{T}{M}\le t \le (MN-1)\frac{T}{M}+T_a$.
\end{lemma}
\begin{IEEEproof}
  Let us first check the periodicity of $u_c(t)$ within the range of $-(M-1)\frac{T}{M}\le t \le (MN-1)\frac{T}{M}+T_a$, which correspond to the start of the first sub-pulse of $u(t+(M-1)\frac{T}{M}$ and the end of the last sub-pulse of $u(t-(M-1)\frac{T}{M})$, respectively.
  From (\ref{ut}), we can divide $u(t)$ into $N$ segments, where $u(t) = \sum_{n=0}^{N-1}u_n(t)$ and the $n$th segment is given by
  $ u_n(t)  =u(t)$ for $nT \le t < (n+1)T$.

  \begin{figure}
    \centering
    \includegraphics[width=8cm]{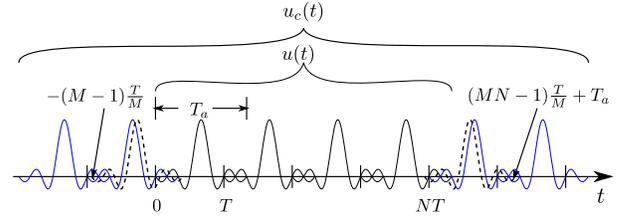}
    \caption{$u_c(t)$ for $D=2$.}
    \label{gce2}
    \vspace{-5mm}
  \end{figure}

  Let $D=\lceil T_a/T \rceil$. If $D=1$, we have
  $
    u_n(t)= a(t-nT),
  $
  which implies that the periodicity within $-(M-1)\frac{T}{M}\le t \le (MN-1)\frac{T}{M}+T_a$ can be obtained by cyclically extending $u(t)$ to
  $u_c(t)=\sum_{n=-1}^{N} a(t-nT)$.
  Similarly, when $D>1$, the periodicity can be obtained by further extending to
  \begin{align}\label{ncpcs}
    u_c(t)=\sum_{n=-D}^{N-1+D} a(t-nT).
  \end{align}
  Two examples of $u_c(t)$ with $D=1,2$ are shown in Fig. \ref{gce1} and Fig. \ref{gce2}, respectively, where the first sub-pulse of $u(t+(M-1)\frac{T}{M})$ and the last sub-pulse of $u(t-(M-1)\frac{T}{M})$ are also plotted with dashed lines.

  Next, let us verify the ambiguity functions. Due to the aforementioned periodicity of  $u_c(t)$, we have
  \begin{align}\label{cycforinte}
    u_c(t)=u_c(t+\dot nT),\,\,\, 0\le \dot n \le N-1,
  \end{align}
  for $m\frac{T}{M} \le  t \le m\frac{T}{M} +T_u$, where $|m| \le M-1$ and $T_u=(N-1)T+T_a$. Then, using (\ref{cycforinte}), the ambiguity function between $u_c(t)$ and $u(t)$ for $|n|\le N-1$ and $|m|\le M-1$ can be calculated similarly to (\ref{northo}), and given by 
  \begin{align}
     & \mathcal A_{u_c,u}(m\frac{T}{M}, n\frac{1}{NT}) \nonumber                                                                 \\
     & =  \int_{m\frac{T}{M}}^{m\frac{T}{M}+T_u} u_c(t)u^*(t-m\frac{T}{M})e^{-j2\pi n\frac{1}{NT}(t-m\frac{T}{M})} dt, \nonumber \\
     & =\delta(n)\delta(m). \label{orthoam}
  \end{align}
  (\ref{orthoam}) completes the proof.
\end{IEEEproof}
Lemma 2 indicates that the constraint of $T_a$ in $u(t)$ can be removed. Once the appropriate CP and CS are added in accordance with (\ref{ncpcs}), the desired local orthogonality can be achieved as well.
As a result, generally the transmit pulse of ODDM modulation is $u_c(t)$, where the extension parameter $D=\lceil T_{a}/T \rceil=\lceil 2Q/M \rceil$. When $M\gg 2Q$, we have $2Q/M\approx 0$. Then, as proved in \cite{oddm}, the ODDM can just employ the DDOP $u(t)$ without cyclic extension ($D=0$).

\section{TF signal localization and numerical results}

\subsection{Frequency domain representation of DDOP}
The frequency domain representation plays an important role in the analysis of  pulse.
In the following, we will derive $U(f)$, the frequency domain representation of $u(t)$.

\begin{figure}
  \centering
  \includegraphics[width=8.8cm]{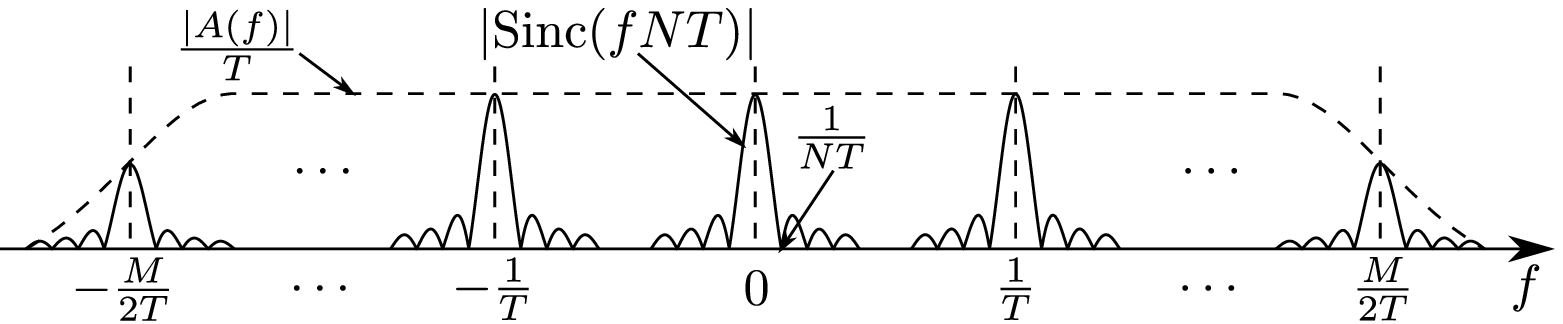}
  \caption{$|U(f)|$.}
  \label{uf}
  \vspace{-5mm}
\end{figure}

It is well-known that the frequency domain representation of an impulse train
$
  \dot u(t)=\sum_{n=-\infty}^{\infty} \delta (t-nT),
$
is a Fourier series and also can be written as an impulse train in frequency domain
$
  \dot U(f)=\frac{1}{T}\sum_{n=-\infty}^{\infty} \delta (f-\frac{n}{T}).
$
It is interesting to observe that the DDOP can be obtained from $\dot u(t)$ by  applying a rectangular window $\rect_{NT}\left(t+\frac{T}{2}\right)$ followed by a $a(t)$ based filtering. Then, we have
\begin{align}
  u\left(t+\frac{T_a}{2}\right)=\left(\dot u(t)\times \rect_{NT}\left(t+\frac{T}{2}\right)\right)\star a(t),
\end{align}
where $\star$ denotes the convolution. Since the multiplication and convolution in time domain correspond to the convolution and multiplication in frequency domain, respectively, we have
\begin{align}
  U(f) & =e^{-j2\pi f \frac{T_a}{2}} A(f) \left(\dot U(f)\star e^{-j2\pi f \frac{(N-1)T}{2}}\sinc(fNT)\right), \nonumber          \\
       & = \frac{e^{-j2\pi f \tilde T}}{T} A(f)  \sum_{n=-\infty}^{\infty} e^{j2\pi \frac{n(N-1)}{2}}\sinc(fNT-nN) , \label{ufeq}
\end{align}
where $\tilde T= (T_a+(N-1)T)/2$ and $A(f)$ is the Fourier transform of $a(t)$. Without loss of generality, let $M$ be an even number. Then, the shape of  $|U(f)|$ in plotted in Fig. \ref{uf}, where the shape of $|\sinc(fNT-nN)|$ is truncated for the purpose of display. Now, it becomes clear that $\sinc(fNT-nN)$ and  $A(f)$ correspond to the orthogonality with respect to $\mathcal F=\frac{1}{NT}$ and $\mathcal T= \frac{T}{M}$, respectively.

\subsection{TF signal localization comparison}
\begin{figure*}
  \centering
  \includegraphics[width=0.96\linewidth]{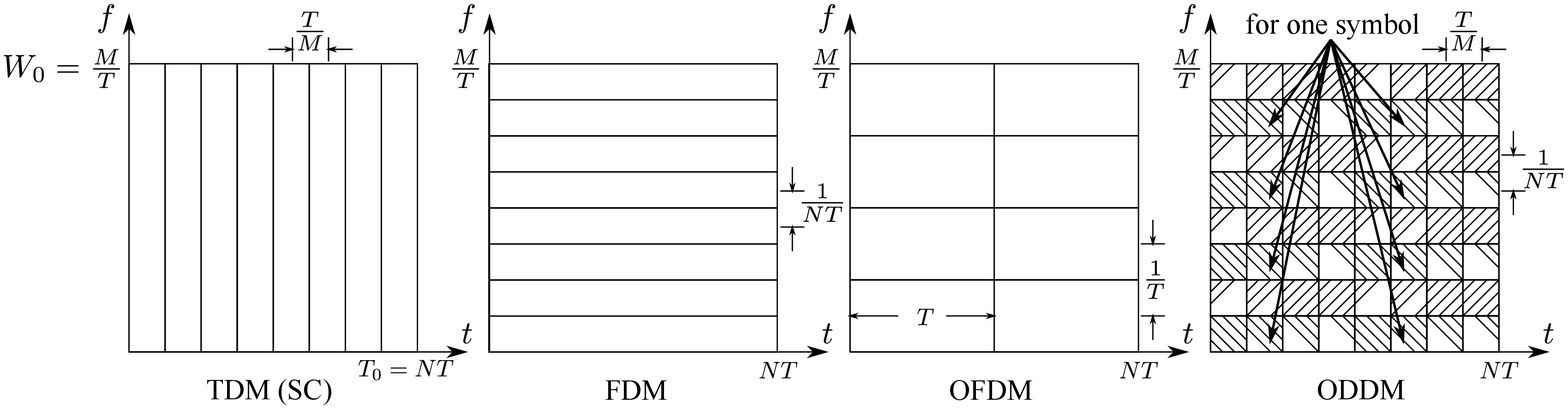}
  \vspace{-2mm}
  \caption{TF signal localization comparison of modulation waveforms.}
  \label{cms}
  \vspace{-5mm}
\end{figure*}
For the TF region bounded by the sampling rate and duration of $W_0=\frac{M}{T}$ and $T_0=NT$, the corresponding degrees of freedom (DoF) of the signal is around $W_0T_0=MN$.
Then, an MC modulation scheme employs $MN$ orthogonal pulses corresponding to its TF resolutions to transmit $MN$ digital symbols, resulting in its own TF localization structure.

With $u(t)$ in (\ref{ut}) and $U(f)$ in (\ref{ufeq}), like that of OFDM in \cite{tff}, the TF signal localization structure of ODDM modulation can be schematically illustrated in Fig. \ref{cms}, where those of other modulation waveforms are also given for comparison. It can be observed that :
\renewcommand{\labelenumi}{\arabic{enumi})}
\begin{enumerate}
  \item For SC modulation, which is a time-division multiplexing (TDM) scheme, the $MN$ digital symbols are conveyed by $MN$ SRN pulses for symbol interval  $\frac{T}{M}$. The pulses are overlapped only in time domain.
  \item For frequency-division multiplexing (FDM) scheme, an example is the OFDM modulation with frequency resolution $\frac{1}{NT}$, where $MN$ digital symbols are conveyed by $MN$ rectangular pulses $\rect_{NT}(t)$ modulated by $MN$ subcarriers, respectively. The pulses are overlapped only in frequency domain.
  \item For the conventional OFDM modulation with frequency resolution $\frac{1}{T}$ and time resolution $T$, $MN$ digital symbols are conveyed by $N$ OFDM symbols, where each OFDM symbols has $M$ rectangular pulses $\rect_{T}(t)$ modulated by $M$ subcarriers, respectively. Since $N$ OFDM symbols are isolated in time domain, these pulses also are overlapped only in frequency domain.
  \item For ODDM modulation with frequency resolution $\frac{1}{NT}$ and time resolution $\frac{T}{M}$, $MN$ digital symbols are conveyed by $M$ pulse trains $u(t)$ modulated by $N$ subcarriers, respectively. These pulses are overlapped in both time and frequency domains to achieve the local orthogonality with respect to $\frac{T}{M}$ and $\frac{1}{NT}$.
\end{enumerate}

\subsection{Numerical results}
\begin{figure}
  \centering
  \includegraphics[width=9cm]{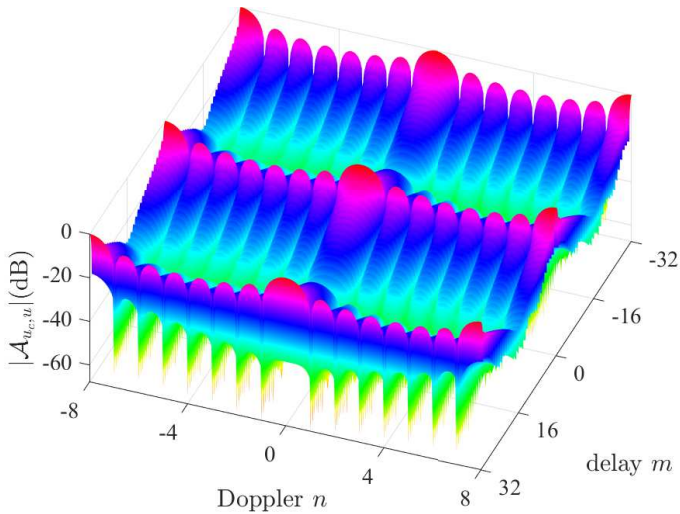}
  \caption{$\left|\mathcal A_{u_c,u}\left(m\frac{T}{M},n\frac{1}{NT}\right)\right|$, $M=32$ and $N=8$.}
  \label{as}
\end{figure}

\begin{figure}
  \centering
  \includegraphics[width=8.8cm]{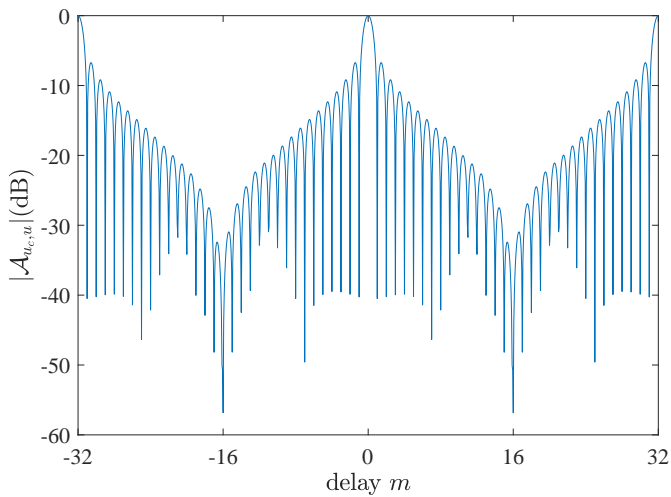}
  \caption{$\left|\mathcal A_{u_c,u}\left(m\frac{T}{M},n\frac{1}{NT}\right)\right|$, $M=32$ and $N=8$, $n=0$.}
  \label{as_m0}
\end{figure}

Now, we present the numerical results for the
ambiguity function of the DDOP.
A three-dimensional plot of the ambiguity function in (\ref{afgceg})
is shown in Fig. \ref{as}, where $\mathcal F=\frac{1}{NT}$, $\mathcal T=\frac{T}{M}$ with $M=32$, $N=8$. $a(t)$ is a root raised cosine pulse with roll-off factor $\rho=0.1$ and $Q=20$. Because $D=2$ for this parameter setting, we adopt the general DDOP design.
The corresponding 2D plot of $\left|\mathcal A_{u_c,u}\left(m\frac{T}{M},n\frac{1}{NT}\right)\right|$ with $n=0$ is also given in Fig. \ref{as_m0}. One can see that with appropriate CP and CS, the DDOP can achieve the local orthogonality within $|m|\le M-1$ and $|n|\le N-1$. For $|m|\ge M$ or $|n|\ge N$, the ambiguity function repeats with time period $T$ and frequency period $\frac{1}{T}$, if we further extend the CP and CS.

The elegant TF localization of ODDM schemes shown in Fig. 6 demonstrates that every information symbol is evenly distributed over its TF region. Thus, it is flexible for allocating TF resources for multi-user communications system design.  In addition, the perfect local orthogonality of the DDOP's ambiguity function with respect to DD resolutions, shown in Figs. \ref{as} and \ref{as_m0}, can be exploited for design integrated sensing and communication (ISAC) systems. We will investigate these topics in our future work.

\section{Conclusion}

In this paper, the recently discovered DDOP is analyzed in terms of local orthogonality, frequency-domain representation and ambiguity function. We clarified the DDOP's local orthogonality and justified its existence as a WH subset, without violating the WH frame theory which governs the global orthogonality corresponding to the WH set. Several sufficient conditions for locally-orthogonal pulses were presented, and a general DDOP design was proposed by introducing CP and CS to the DDOP. We derived the DDOP's frequency domain representation, and compared the DDOP-based ODDM modulation with other modulation schemes, in terms of TF signal localization. We demonstrated the perfect local orthogonality of DDOP with respect to DD resolutions by its ambiguity function.


%





\ifCLASSOPTIONcaptionsoff
  \newpage
\fi



%
\vspace{-3mm}
\bibliographystyle{IEEEtran}
\bibliography{oddm}




%








\end{document}